\documentclass[10pt]{article} 
\usepackage{epsfig, graphicx}
\usepackage[multiple]{footmisc}
\usepackage{color,cite}
\usepackage{hyperref, url}
\usepackage{amsmath}
\usepackage[textwidth=7.2in,textheight=9.7in]{geometry}
\newcommand{\fig}[1]{Fig.\ \ref{#1}}
\def\eqn#1{eq.\ (\ref{#1})}

\def\bea{\begin{eqnarray}}
\def\eea{\end{eqnarray}}
\def\nn{\nonumber}

\def\be{\begin{equation}}
\def\ee{\end{equation}}
\def\barr{\begin{array}}
\def\earr{\end{array}}
\def\bfig{\begin{figure}}
\def\efig{\end{figure}}

\def\eqn#1{eq.\ (\ref{#1})}
\def\fig#1{fig.\ (\ref{#1})}

\newcommand{\diag}{\mathrm{diag}}

\newcommand{\Neff}{N_\mathrm{eff}}

\newcommand{\Hbar}{\widetilde{H}}
\newcommand{\Ebar}{\widetilde{\cal E}}
\newcommand{\tep}{\tilde{\varepsilon}}
\newcommand{\ffd}{f_{\rm FD}}

\newcommand{\nns}{n_{\nu_s}}

\begin{document}
\title{\bf Inflation models in the light of self-interacting sterile neutrinos}
\author{Arindam Mazumdar$^a$\footnote{e-mail: \tt arindam.mazumdar@iitkgp.ac.in},
Subhendra Mohanty$^b$\footnote{e-mail: \tt mohanty@prl.res.in}, 
Priyank Parashari$^{b,c}$\footnote{e-mail: \tt parashari@prl.res.in}
}
\date{}
\maketitle
\vspace*{-1cm}
\begin{center}
{‌\small $a$) Centre for Theoretical Studies, Indian Institute of Technology, Kharagpur -721302, India\\
$b$) Physical Research Laboratory, Ahmedabad- 380009, India  \\
$c$) Indian Institute of Technology, Gandhinagar, 382355, India }
\end{center}

\begin{abstract}
{ Short baseline neutrino experiments, like LSND and MiniBooNE experiments, pointed towards the existence of eV mass scale sterile neutrinos. To reconcile sterile neutrinos with cosmology self interaction between sterile neutrinos has been studied. 
We analysed Planck cosmic microwave background (CMB) data with self-interacting 
sterile neutrino (SI$\nu$) and study their impact on inflation models. The fit 
to the CMB data in SI$\nu$ model is as good as the fit to $\Lambda$CDM model. We 
find that the spectral index ($n_s$) values shift to $0.9361\pm 0.0055$ in 
SI$\nu$ model. This has significant impact on the validity of different 
inflation models. For example the Starobinsky and quartic hilltop model, which 
were allowed within $\Lambda$CDM cosmology, are ruled out. On the other hand 
some models like natural and Coleman-Weinberg inflation are now favoured. 
Therefore, the existence of self interacting sterile neutrinos with eV order of 
mass will play an important role in the selection of correct inflation model.}

\end{abstract}
\maketitle
\section{Introduction}
\label{sec:intro}
Some short-baseline (SBL) experiments have reported anomalies in the neutrino events which cannot be explained within the framework of standard 3 active neutrinos oscillation. This was pointed out by the LSND experiment~\cite{Aguilar:2001ty}, where an excess of electron like events was reported in $\bar{\nu}_\mu\rightarrow \bar{\nu}_e$ oscillations. Similar kind of excess has also been reported in both the neutrino and anti-neutrino channels by the
MiniBoone experiment recently~\cite{AguilarArevalo:2008rc,AguilarArevalo:2010wv,Aguilar-Arevalo:2018gpe}. There are a few other experiments which have also reported similar anomalies~\cite{Anselmann:1994ar,Hampel:1997fc,Mention:2011rk}. To explain these anomalies in terms of neutrino oscillations, at least one additional neutrino species is required. This extra neutrino species must be sterile to be compatible with the constraint on number of light neutrinos coupled to $Z$ boson~\cite{ALEPH:2005ab}. These observations in SBL experiments have motivated many authors in recent years to study the 3+1 neutrino scenario~\cite{Kopp:2011qd,Giunti:2011hn,Conrad:2012qt,Kopp:2013vaa,Gariazzo:2013gua,Collin:2016rao,Dentler:2018sju,Denton:2018dqq,Chauhan:2018dkd}. However, there has been a long standing problem of accommodating light sterile neutrino in the cosmological context~\cite{Hamann:2011ge,Hannestad:2012ky}. The problem with light sterile neutrino is that their allowed mass and mixing angles from neutrino experiments will lead to their complete thermalization in the very early universe. Therefore they will contribute to radiation energy density which will result in conflict with the existing bounds on number of relativistic degrees of freedom obtained from Big Bang Nucleosynthesis (BBN)~\cite{Cyburt:2015mya} and 
Cosmic Microwave Background (CMB) data~\cite{Ade:2015xua,Aghanim:2018eyx}. Also fully thermalized sterile neutrinos of mass ${\cal O}({\rm eV})$ will lead to suppression in the structure formation resulting in a tension with allowed neutrino mass bounds obtained from cosmological observations~\cite{Hannestad:2010kz,Aghanim:2018eyx, Wong:2011ip,Hamann:2011ge,Lesgourgues:2012uu}. Hence, in order to make light sterile neutrinos viable with cosmological observations, some new physics is required. Many solutions were proposed to accommodate the light sterile neutrinos in the cosmology~\cite{Giusarma:2011zq,Ho:2012br,Gelmini:2004ah,Foot:1995bm,Bento:2001xi}. 
Recently, it was proposed that by introducing self interaction in the sterile neutrino sector this problem can be resolved~\cite{Hannestad:2013ana,Dasgupta:2013zpn,Chu:2018gxk}. We will call this model as the SI$\nu$ model. In this paper we study in detail the consequences of introducing secret interaction in 
sterile neutrino sector on the models of inflation.
 
It has been pointed out that neutrino self interaction with coupling $\sim 10^{10}G_F$ can prevent oscillations of sterile neutrinos into active ones till sub-MeV temperatures and restrict the contribution to $N_{\rm eff}$ within the range allowed by observations~\cite{Hannestad:2013ana,Dasgupta:2013zpn,Chu:2018gxk,Song:2018zyl}. Also large self-interaction prevents the free-streaming of the sterile neutrinos till the time they become non-relativistic, thereby preventing them from erasing small scale structures~\cite{Mirizzi:2014ama,Chu:2015ipa}.

Introduction of self interaction in the sterile neutrino sector leads to significant changes in the cosmological parameters allowed from CMB observations\cite{Forastieri:2017oma,Song:2018zyl}. In these papers fits of both cosmological and neutrino parameters were done with the CMB observations.
However, in the present paper we take the best fit values of mass and mixing angles for three active neutrinos from the global fit of the neutrino oscillation data~\cite{Esteban:2018azc}, and for sterile neutrinos from the MiniBooNE experiments~\cite{Aguilar-Arevalo:2018gpe}. 
Strength of self interaction $\sim 10^{10}G_F$, which evades the bounds from BBN and LSS surveys, was kept fixed and seven standard cosmological parameters were varied to get the best fit for the CMB data. By following this procedure
the most important change we find is the lowering of spectral index ($n_s$). The Planck CMB data had narrowed down the $n_s$-$r$ allowed region in the $\Lambda$CDM model~\cite{Ade:2015lrj,Akrami:2018odb}, and had disfavoured many inflation models like natural inflation and Coleman-Weinberg inflation.
However, some models like
Starobinsky inflation was in very good agreement with the Planck data in the $\Lambda$CDM cosmology~\cite{Ade:2015lrj,Akrami:2018odb}. 
The Starobinsky model predicts spectral index 
in a very narrow range and any shift in the prediction is not possible by changing the model parameters. In our analysis we will
show later that self-interacting sterile neutrinos shift the spectral index in such a low value that Starobinsky inflation  cannot be accommodated by the CMB data. However, we find that other inflation models like natural inflation and Coleman-Weinberg inflation are now favored in SI$\nu$ model. The effect of self-interaction in the active neutrino sector on CMB was considered in ref~\cite{Kreisch:2019yzn} and its effect on constraining 
inflation models was discussed in ref~\cite{Barenboim:2019tux}. The effect of additional relativistic degrees of freedom on selection of inflation models were
considred in ref.~\cite{Tram:2016rcw}.
However, in these studies the range of $n_s$ increases and allows a larger set of inflation
models to fit in, which is quite contrary to our case.

The paper is organised as follows. In section~\ref{sec:sinu} we have reviewed the self-interaction in the sterile neutrino sector and how 
it affects the thermalization of sterile neutrinos. In section~\ref{sec:pert} we have discussed  the cosmological perturbation 
equations in the SI$\nu$ model. In  section~\ref{sec:cmb} we study the effect of self-interaction
and mass of the sterile neutrino on the CMB and also discuss the effect on cosmological parameters by including the perturbation equations with self interaction in the CLASS code and doing a Markov Chain Monte Carlo (MCMC) analysis by Montepython. In 
section~\ref{sec:inf} we discuss inflation models which are now favored or disfavoured in SI$\nu$ model. We sum up our work and give our conclusions
in section~\ref{sec:conclusion}.
\section{Self-interacting sterile neutrino in cosmology}\label{sec:sinu} 
It is well established from the results of neutrino oscillation experiments that neutrinos oscillate among flavor eigenstates, which is caused by non-zero neutrino mass and mixing~\cite{Fukuda:1998mi,Ahmad:2001an,GonzalezGarcia:2007ib}. However, some SBL experiments, like LSND and MiniBooNE experiments, have reported anomalies in the standard three active flavours neutrino model.  
In order to explain these observations by neutrino oscillation, we need $\Delta m^2\sim {\cal O}({\rm eV}^2)$ and large mixing between the active and sterile neutrinos~\cite{Kopp:2011qd,Giunti:2011hn,Conrad:2012qt,Kopp:2013vaa,Gariazzo:2013gua,Collin:2016rao,Dentler:2018sju,Denton:2018dqq}. Therefore these anomalies indicates the possibility of additional neutrino species which can not interact with standard model particles. In this work we will consider one extra sterile neutrino.
 As neutrinos now oscillate among 4 different flavours, the flavour eigenstates, denoted by Greek letters($\alpha, \beta$), are related to the mass eigenstates, denoted with Roman letters($i,j..$) , as
\begin{equation}
 |\nu_\alpha\rangle = U_{\alpha i}^\dagger |\nu_i\rangle ,
\end{equation}
where $\alpha=(e,\mu,\tau,s)$, $i$ runs from 1 to 4 and $U$ is the neutrino mixing matrix.
However accommodating one light sterile neutrino in cosmology has been a long standing problem. If the mixing between active and sterile neutrinos is sufficiently large, they will come in thermal equlibrium with the active neutrinos via oscillation. Therefore, light sterile neutrinos will contribute to effective number of neutrino species $N_{\rm eff}$, which is very well constrained by measurement of primordial abundances produced in BBN~\cite{Cyburt:2015mya} and CMB data~\cite{Ade:2015xua,Aghanim:2018eyx}. 
In 3+1 neutrinos scenario with sterile neutrinos of mass ${\cal O}({\rm eV})$ and non-zero mixing angle, effective number of neutrino species 
$N_{\rm eff}$ is equal to 4 at MeV temperatures, which is in tension with the current constraint obtained from the BBN and CMB data .
 In addition to these constraints, Large Scale Structure (LSS) observations also put bound on neutrinos mass. This mass bound obtained from LSS observations is also in tension with the eV order sterile neutrinos~\cite{Hannestad:2010kz,Aghanim:2018eyx, Wong:2011ip,Hamann:2011ge,Lesgourgues:2012uu}. Hence some new physics needs to be invoked in order to evade these constraints.

Self interaction between sterile neutrinos has been proposed in order to alleviate this tension~\cite{Hannestad:2013ana,Dasgupta:2013zpn}. 
There are two different types of self-interactions which are usually considered. One is pseudo-scalar mediated interaction~\cite{Archidiacono:2014nda}
and another invokes a gauge interaction~\cite{Hannestad:2013ana,Dasgupta:2013zpn}. The interaction term in the Lagrangian for the pseudo-scalar
looks like
\be
\mathcal{L}_{\rm int}= g_\Phi \bar{\nu}_s\Phi\gamma_5\nu_s\, ,
\ee
whereas, in the case of interaction by gauge-boson it takes the following form.
\be
\mathcal{L}_{\rm int}= g_X \bar{\nu}_s\gamma_\mu\frac{1}{2}(1-\gamma_5)\nu_sX^\mu\,.
\label{eq:Lagrang}
\ee
Here $g_\Phi$ and $g_X$ are Yukawa coupling and gauge coupling respectively. On the energy scales smaller than
the mediator mass, We can effectively write these two kind of interactions as four-Fermi interaction where the 
effective interaction strength between four neutrinos reduces to $\frac{G_\Phi}{\sqrt{2}}=\frac{g_\Phi^2}{8M_\Phi^2}$ and
$\frac{G_X}{\sqrt{2}}=\frac{g_X^2}{8M_X^2}$. In this paper we consider the interaction mediated by gauge boson only.
Initially the $N_{\rm eff}$ is shared between the active neutrinos and strong self-interaction in the sterile sector 
prevents the sterile neutrinos to be thermalized with the active neutrinos by oscillation. Later on the sterile neutrinos gets thermalized but still its contribution to $N_{\rm eff}$ is smaller than that of active neutrinos and the total $N_{\rm eff}$ is $\sim 3$.
 The self interaction between the sterile neutrinos induces scattering among them in the early universe and this scattering rate is given as
\begin{equation}
\Gamma_X=\nns\langle \sigma v\rangle \simeq G_X^2T_{\nu_s}^5 \;,
\end{equation}
where $\nns$ and $T_{\nu_s}$ are the number density and temperature of
the sterile neutrinos. 
Massive neutrinos affect the background as well as cosmological perturbation evolutions. As neutrinos interact very weakly, they free stream within cosmological plasma. If neutrinos free-stream till the time they become non-relativistic, they will give rise to a suppression in the growth of perturbations on smaller scales. On the other hand if there is a large self interaction within sterile neutrino sector, it can delay the free-stream regime till the epoch self interaction rate becomes smaller than the Hubble rate. Therefore, if $G_X$ is sufficiently large, the free-stream regime can be delayed till the time neutrinos turn non-relativistic. In that case, sterile neutrinos always scatter via self interaction and they will never have a free-stream regime~\cite{Mirizzi:2014ama,Chu:2015ipa}. We can calculate the smallest $G_X$ for which this happens by equating the Hubble rate ($H$) with the interaction rate
\be
G_X^2 T_{\nu_s}^5 \approx H(T_{\gamma}).
\ee
Solving this for the Hubble rate in eV range and neutrino temperature, we get the $G_X \sim 10^{10} G_F$. Therefore, if coupling $G_X$ is large or equal to $10^{10} G_F$, mass bound obtained from LSS is not applicable in that case (for detailed discussion see ref.~\cite{Mirizzi:2014ama,Chu:2015ipa}). Hence, we take $G_X = 10^{10} G_F$ in order to evade the mass bound from the LSS observations. 

In order to study quantitatively the effect of self interaction in sterile sector we will calculate the time evolution of density matrix ($\rho$) for all the neutrino flavors. The density matrix in the two bases, mass and flavour, are connected as
\begin{eqnarray}
 \rho_{\alpha\beta} = U_{\alpha i}^\dagger \rho_{ij} U_{j \beta} ~.
\end{eqnarray}
We therefore solve the quantum kinetic equations (QKEs) of the 3+1 neutrino
ensemble as described in Ref.~\cite{Mirizzi:2012we}. QKEs are given as
\begin{equation}
{\rm i}\,\frac{d\varrho}{d t} =[\Omega,\varrho]+ C[\varrho]\, .
\label{eq:drhodt}
\end{equation}
where $C[\varrho]$ is the collision term and 
\begin{equation}
\Omega=\frac{1}{2p}U^\dagger
M^2U+\sqrt{2}G_F
\left[-\frac{8p}{3}\left(\frac{{\cal E}_l}  {M_W^2}
  +\frac{{\cal E}_\nu}{M_Z^2}\right)\right]
+\sqrt{2}G_X\left[-\frac{8p}{3}\frac{{\cal E}_s}{M_X^2}\right]\,.
\label{eq:matterpotential}
\end{equation}
The first term represents the oscillation between the flavor and mass eigenstates. Here $U=R_{34}R_{24}R_{23}R_{14}R_{13}R_{12}$
 is the neutrino mixing matrix and $M$ is the mass matrix whose components are determined from the oscillation experiments. Here $R_{ij}$ is the rotation matrix corresponding to angle $\theta_{ij}$ in the
${ij}$ plane.
Here in this work we have used $M^2={\rm diag} (0,\Delta m_{21}^2,\Delta m_{31}^2, m_{\rm st}^2)$. We have used the latest result for the standard three active neutrino oscillation parmeters obtained from global analysis of neutrino data~\cite{Esteban:2018azc}. We have chosen the best fit values for $\Delta m^2_{21}=7.39\times10^{-5} \rm{eV^2}$, $\Delta m_{31}^2=2.523\times10^{-5} \rm{eV^2}$, $\theta_{12}=33.82^{\circ}$, $\theta_{31}=8.61^{\circ}$ and  $\theta_{23}=48.3^{\circ}$\cite{Esteban:2018azc}. Whereas mass of the sterile neutrino $m_{\rm st} = 0.2\rm{eV}$ and mixing angle $\sin^2\theta_{14}=0.95$,  which are the best fit value obtained from MiniBooNE experiments~\cite{Aguilar-Arevalo:2018gpe}.
The second term in \eqn{eq:matterpotential} represents the electro-week interaction between the neutrinos and 
the electrons that are present in the baryonic 
fluid after big bang nucleosynthesis(BBN). In this term ${\cal E}_l$, ${\cal E}_\nu$ and ${\cal E}_s$ are the energy density for electrons, active and sterile neutrinos. Here $M_W$ and $M_Z$ are the masses of $W$ and $Z$ bosons respectively.
The third in \eqn{eq:matterpotential} purely corresponds to the self interaction between the sterile neutrinos.

\begin{figure}
\begin{center}
 \includegraphics[width=10cm,height=8cm]{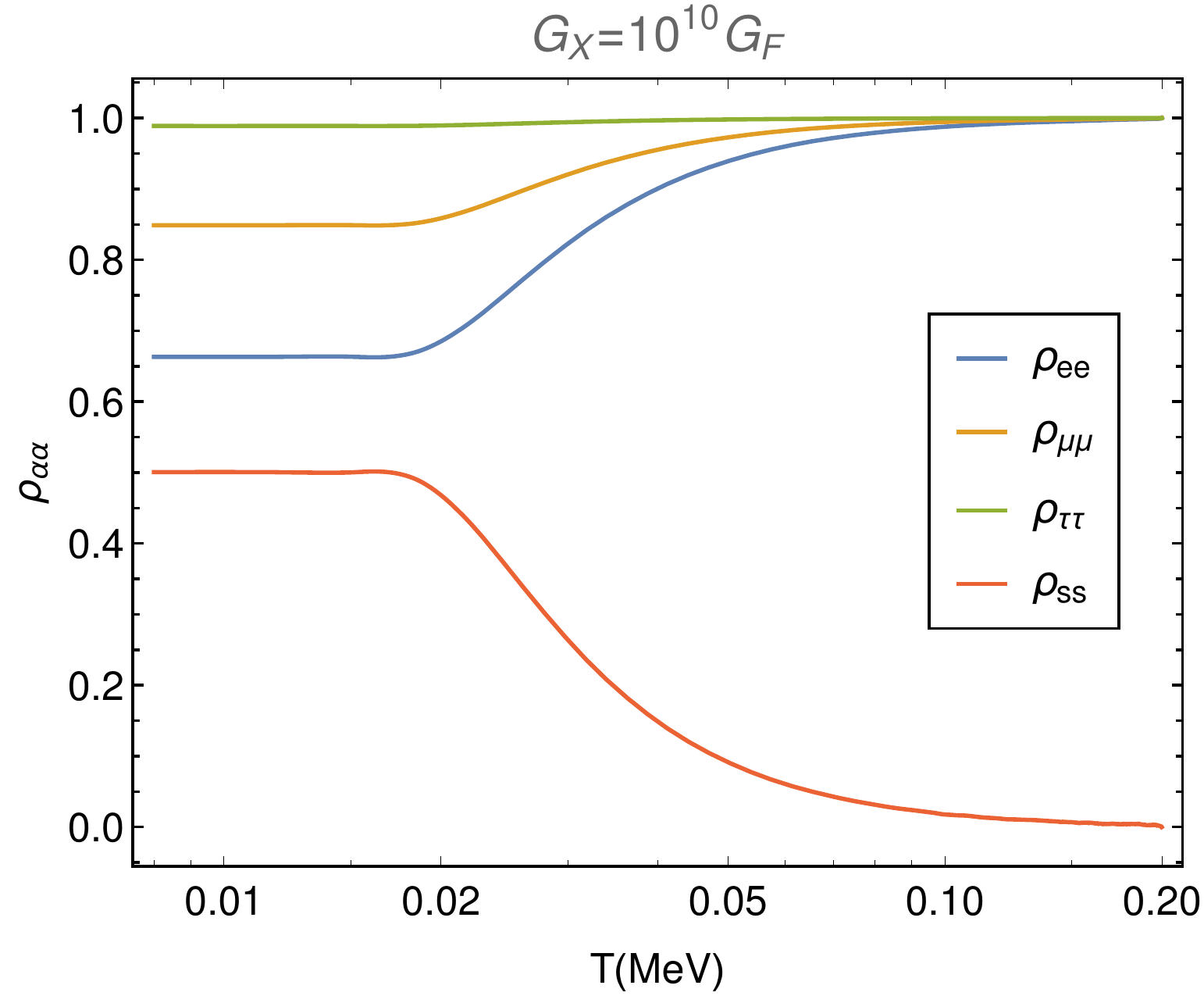}
 \caption{Self interaction delays the thermal mixing of neutrinos.}\label{fig:qke}
 \end{center}
 \end{figure}

QKEs as given in~\eqn{eq:drhodt} are very computationally demanding since density matrix has 
momentum dependence. In order to reduce the computation time and also to retain the important features of the time evolution, we have worked in the average momentum approximation as described
in~\cite{Mirizzi:2012we}. Under this approximation, we assume
\begin{equation}
  \varrho(x,y)\longrightarrow \ffd(y)\rho(x) \;,
\label{eq:rhoav}  
\end{equation}
where $\ffd$ is the Fermi-Dirac distribution function, $x= ma$ and $y= pa$. We have also introduced new variables for convenience as
\begin{equation}
\quad z_\gamma\equiv T_\gamma a,\quad
z_\nu\equiv T_\nu a,\label{eq:param}  
\end{equation}
where we have chosen the  mass scale $m$ = 1MeV. We put~\eqn{eq:rhoav} and~\eqn{eq:matterpotential} in~\eqn{eq:drhodt} and made the change of variables as given in~\eqn{eq:param} to get
\begin{eqnarray}\label{eq:qke}
i\frac{d\rho}{d x}=&+&\frac{x^2}{2m^2\Hbar}\langle
\frac{1}{y}\rangle\left[U^\dagger M^2
  U,\rho\right]+\frac{\sqrt{2}G_F m^2}{x^2
  \Hbar}\left[\left(-\frac{8 \langle y\rangle
    m^2}{3x^2M_W^2}\Ebar_l-\frac{8 \langle y\rangle
    m^2}{3x^2M_Z^2}\Ebar_\nu\right),\rho\right]\nonumber\\ &+&\frac{\sqrt{2}G_X
  m^2}{x^2 \Hbar}\left[-\frac{8 \langle y\rangle
    m^2}{3x^2M_X^2}\Ebar_s,\rho\right]+\frac{x\langle
  C[\rho]\rangle}{m\Hbar}\,,
\label{eq:avedensitymatrix}
\end{eqnarray}
where
\begin{eqnarray}
  \Ebar_l&=&\diag  (\tep_e,0,0,0)\,,\\
  \Ebar_\nu&=&\frac{2}{2\pi^2}\int_0^\infty
d y y^3
G_s\varrho(x,y)G_s=\frac{7}{8}\frac{\pi^2}{15}G_s\rho(x)G_s\,,\\
\Ebar_s&=&\frac{2}{2\pi^2}\int_0^\infty
d y y^3
G_{sX}\varrho(x,y)G_{sX}=\frac{7}{8}\frac{\pi^2}{15}G_{sX}\rho(x)G_{sX}\,,
\end{eqnarray}
 where $G_s=\diag(1,1,1,0)$ and $G_{sX}=\diag(0,0,0,1)$. Also here $\Hbar$ is the normalized Hubble parameter given as
\begin{equation}
\Hbar\equiv \frac{x^2}{m}H=\frac{m}{M_{\rm
    Pl}}\sqrt{\frac{\tep(x,z_\gamma(x),z_\nu(x))}{3}}\, .
\end{equation}
In this expression $M_{\rm Pl}$ and $\tep\equiv \varepsilon(x/m)^4$ are the reduced Planck mass and the comoving total energy respectively, where
$\varepsilon=\tep_e+{\rm Tr}(\Ebar_\nu+\Ebar_s)$ is  the total energy
density.

Now the collision term $\langle C[\varrho]\rangle$ under average momentum approximation  is given
as~\cite{Chu:2006ua,Mirizzi:2012we}
\begin{eqnarray}
\langle C_\nu[\varrho]\rangle &=&
-\frac{i}{2}G_F^2\frac{m^5}{x^5}(\{S^2,\rho-I\}-2S(\rho-I)S+\{A^2,\rho-I\}+2A(\rho-I)A\,,\nonumber\\ \langle
C_s[\varrho]\rangle &=&
-\frac{i}{2}G_X^2\frac{m^5}{x^5}(\{S_X^2,\rho-I\}-2S_X(\rho-I)S_X)\,,
\end{eqnarray} 
where $S$ and $A$ are the active neutrino scattering and annihilation matrix which are given as
$S=\diag(g_s^e,g_s^\mu,g_s^\tau,0)$ and
$A=\diag(g_a^e,g_a^\mu,g_a^\tau,0)$. Here
$(g_s^e)^2=3.06$, $(g_a^e)^2=0.50$, $(g_s^{\mu,\tau})^2=2.22$,
and $(g_a^{\mu,\tau})^2=0.28$ ~\cite{Chu:2006ua}. And for sterile neutrinos $S_X=\diag(0,0,0,1)$. 
Next, we calculate the value of $N_{\rm eff}$ which is just the trace of the density matrix and is given as
\begin{equation}
\Neff={\rm Tr}\rho(T)=\sum_{\alpha=1}^{4}\rho_{\alpha\alpha}. 
\label{eq:Neffgood}
\end{equation}    

Then we solve the QKEs for the $3+1$ neutrino scenario with self interaction in the sterile neutrino sector. We have plotted number density $\rho_{\alpha\alpha}$ against temperature in \fig{fig:qke}. It is clear from the \fig{fig:qke} that
 strong self interaction within sterile neutrinos, $G_X=10^{10}G_F$, can prevent the mixing between active and sterile neutrinos till sub MeV scales. Moreover, after sterile neutrinos come into equilibrium with active neutrinos, all four neutrinos do not contribute equally to the $N_{\rm eff}$. They redistribute themselves such that $N_{\rm eff}$ always remains close to 3 because of the self interaction, which is in the allowed range obtained from the BBN and CMB data. Therefore, self interaction successfully evades all cosmological constraints.
\section{Perturbation equations}
\label{sec:pert}
In this section we briefly describe the perturbation theory in the presence of self interacting sterile neutrinos. 
Perturbation equations for neutrinos are solved in mass eigenstates. Therefore the density matrix in mass eigen basis 
represents the distribution function for the the neutrinos. So, we write
\begin{eqnarray}
 f_i = f_{\rm FD}\sum_\alpha|U_{\alpha i}|^2 \rho_{\alpha \alpha}.
\end{eqnarray}
Naturally this distribution function is not normalized to one. However,
for the purpose of calculating perturbed Boltzmann equation, normalization of the distribution function is not necessary.
$\sum_\alpha|U_{\alpha i}|^2 \rho_{\alpha \alpha}$ has to be evaluated from the QKE and it depends on time. 
We perturb this distribution function as,\cite{Ma:1995ey} 
\begin{eqnarray}\label{eq:pdf-pert}
 f_i = f_0|U_{\alpha i}|^2 \rho_{\alpha \alpha} (1+\Psi(x^i,q,n_j,\tau)) = \bar{f_0}(1+\Psi(x^i,q,n_j,\tau)),
\end{eqnarray}
where $f_0$ is the zeroth order Fermi-Dirac distribution function. Here $x^i$ denotes the spatial coordinates, $\tau$ is the 
conformal time and $q$ is the amplitude of momentum divided by the energy (see ref~\cite{Ma:1995ey}). $n_i$ are the directions 
of the momentum vectors.
The phase space distribution evolves according to the Boltzmann equation
\begin{equation}\label{eqn:boltzmann}
	{Df_i \over d\tau} = {\partial f_i \over \partial \tau}
	+ {dx^i \over d\tau}{\partial f_i\over \partial x^i}
	+ {dq \over d\tau}{\partial f_i\over \partial q}
	+ {dn_i \over d\tau}{\partial f_i\over \partial n_i}
	= \left( {\partial f_i \over \partial\tau} \right)_C\,.
\end{equation}
Since $\sum_\alpha|U_{\alpha i}|^2 \rho_{\alpha \alpha}$ has time dependence, ${\partial f_i \over \partial \tau}$ will 
introduce a time derivative of $\sum_\alpha|U_{\alpha i}|^2 \rho_{\alpha \alpha}$ in the 
Boltzmann equation. From \fig{fig:qke} we see that the changes in $\rho_{\alpha \alpha}$ occurs at a temperature above $10^{-2}$MeV. 
However, the Boltzmann codes like CAMB or CLASS starts solving the perturbation equations from much lower energy scales.
Therefore, the $\sum_\alpha|U_{\alpha i}|^2 \rho_{\alpha \alpha}$ remains constant and time derivative does not appear
in the Boltzmann hierarchy equation. 

The term on the right hand side of \eqn{eqn:boltzmann} is the collision term.
This term is zero in the standard three neutrino case. In relaxation time approximation for the self interacting neutrinos this term is 
written as~\cite{Hannestad:2000gt}
\be
\frac{1}{\bar{f_0}}\frac{\partial f_i}{\partial \tau}=-\frac{\Psi}{\tau_\nu}\,, \nn 
\ee 
where 
\be
\tau_\nu^{-1}={a}\nns\langle\sigma v\rangle=\frac{3}{2}\frac{\zeta(3)}{\pi^2}{a}G_X^2 {T_\nu^5 \rho_{ss}}\,.\nn
\ee
Here, the value of $\rho_{ss}$ is taken from the solution of QKEs~\eqn{eq:qke}. 
Using this collision term we can rewrite the Boltzmann equation as\cite{Ma:1995ey}
\begin{equation}
\frac{\partial \Psi_i}{\partial \tau} + i \frac{q(\vec{k} \cdot \hat{n})}{\epsilon} \Psi_i + \frac{d \ln f_0}{d \ln q} \left[ \dot{\eta} - \frac{\dot{h}+6\dot{\eta}}{2} \left(\hat{k} \cdot \hat{n} \right)^2 \right] = - \Gamma_{ij} \Psi_j\, ,
\end{equation}
where $\Gamma_{ij} =  U{\rm diag} (0,0,0,1) U^\dagger/\tau_\nu $.
We expand $\Psi$ in a Legendre series
as
\be
\Psi(\vec k,\hat n, q,\tau) = \sum_{\ell =0}^{\infty} (-i)^\ell (2\ell +1)\,\Psi_\ell (\vec k, q,\tau) P_\ell(\hat k . \hat n)\, .
\ee
Using this series expansion we write the Boltzmann equation as a hierarchy of multipoles.
\begin{subequations}\label{eq:scalar-pert}
\begin{align}
&\dot{\Psi}_{i,0} = -\frac{q k}{\epsilon} \Psi_{i,1} + \frac{1}{6} \dot{h} \frac{d \ln{f_0}}{d \ln{q}} \, ,\\
&\dot{\Psi}_{i,1} = \frac{qk}{3 \epsilon} \left(\Psi_{i,0} - 2 \Psi_{i,2} \right) \, ,\\ 
&\dot{\Psi}_{i,2} = \frac{qk}{5 \epsilon} \left( 2 \Psi_{i,1}- 3 \Psi_{i,3} \right) - \left( \frac{1}{15} \dot{h} + \frac{2}{5}\dot{\eta} \right) \frac{d \ln{f_0}}{d \ln{q}} - \Gamma_{ij} \Psi_{j,2}\, ,\\
&\dot{\Psi}_{i,\ell} = \frac{qk}{(2\ell+1)\epsilon} \Big[ \ell \Psi_{i,(\ell-1)} - (\ell+1) \Psi_{i,(\ell+1)} \Big]  - \Gamma_{ij} \Psi_{j,\ell} \quad (\ell\ge 3)\, ,
\end{align}
\label{eq:boltz_mass_int}
\end{subequations}
Following ref~\cite{Forastieri:2017oma} we have kept the collision term zero in the $\ell=0$ and $\ell=1$ equations for conserving particle number  and momentum. 
We have modified the Boltzmann code CLASS~\cite{Lesgourgues:2011re,Blas:2011rf,Lesgourgues:2011rh} with the above mentioned equations.

Similarly, as we have introduced scalar perturbations in \eqn{eq:pdf-pert} we can introduce tensor perturbations in the distribution function.
\begin{eqnarray}
  f_i = f_0|U_{\alpha i}|^2 \rho_{\alpha \alpha} (1+n_i n_je_{ij}\Delta^T(x^i,q,n_j,\tau))\,
\end{eqnarray}
where $e_{ij}$ is a traceless-transverse tensor and $\Delta^T$ is the neutrino tensor transfer function. Using the similar procedure like scalar perturbation we can come up with a similar set of equations for $\Delta^T$ like \eqn{eq:scalar-pert}.
These equations can be found in ref~\cite{Ghosh:2017jdy}. We have changed those equations in CLASS and could not find any visible effect
on $BB$ mode spectrum of CMB.
\section{Effect on CMB}\label{sec:cmb}
\begin{figure}
\centering
\includegraphics[width=9.cm]{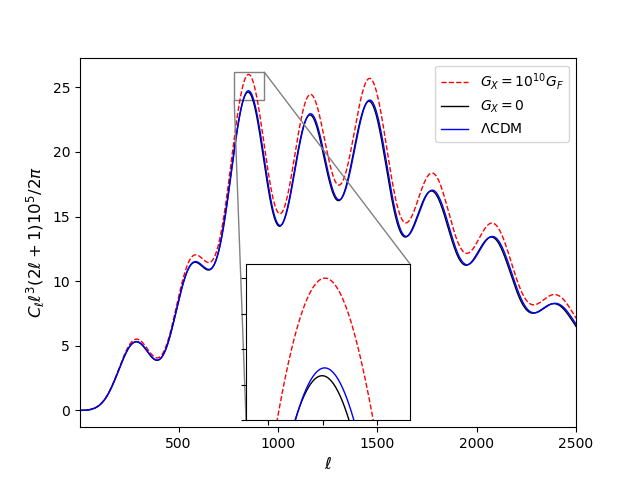}
 \includegraphics[width=9.cm,height=7cm]{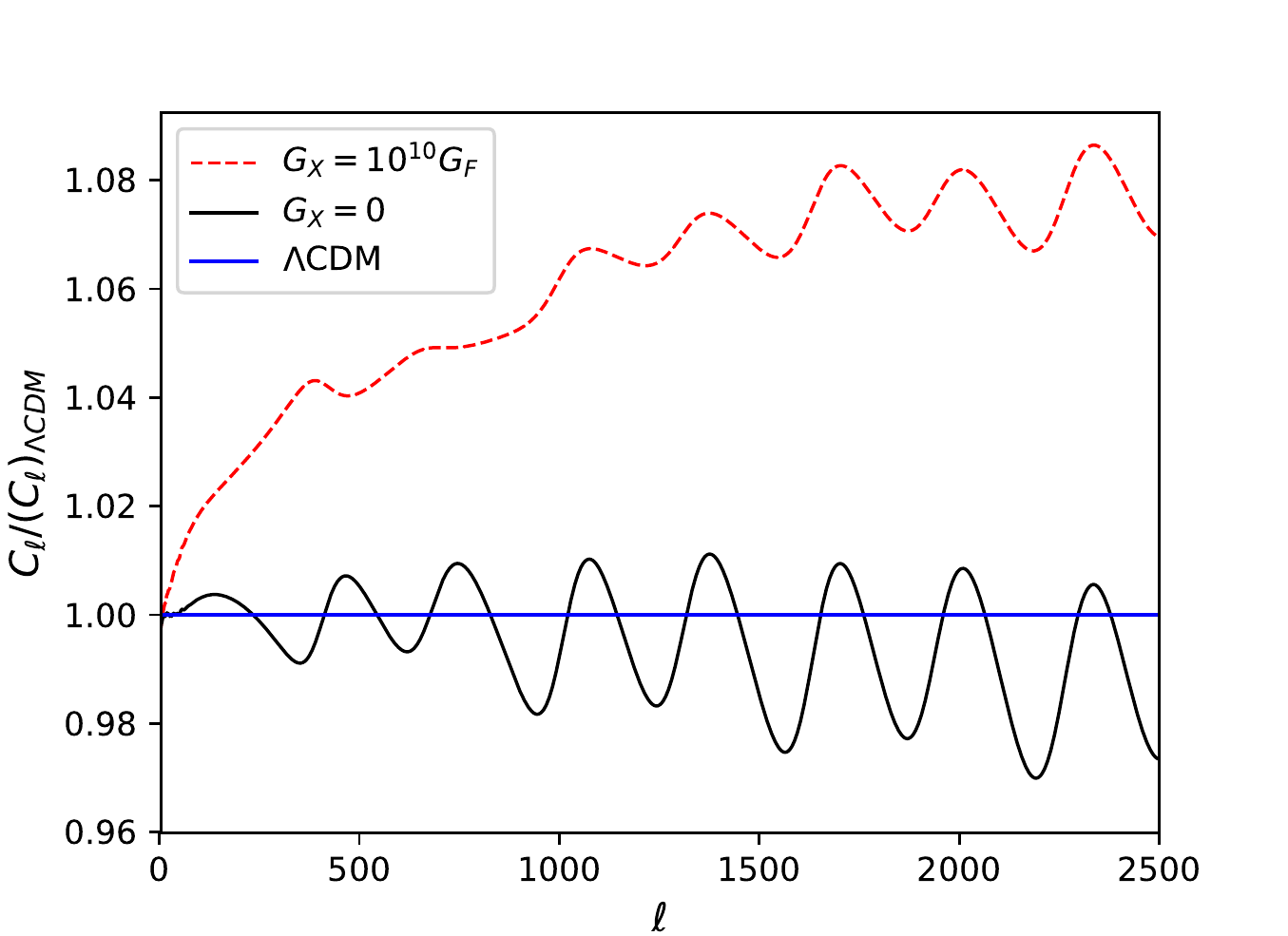}
 (a)\hspace*{9cm} (b)
\caption{$\Lambda$CDM model accommodates three massive neutrinos in normal 
hierarchy. Sterile neutrino with no interaction takes
 the masses and degeneracy factors given in table~\ref{tab:degen}. The red 
dashed line shows the effect of self interaction in sterile neutrino sector.}
 \label{fig:cmb}
 \end{figure}

We would compare two models here in this paper. The first model is $\Lambda$CDM with three massive neutrinos which has  masses 
corresponding to normal hierarchy and lowest neutrino mass to be $10^{-5}$eV.  Such a small mass for the lowest massive neutrino makes it 
effectively massless. Here all the three neutrinos have degeneracy equal to one.
We will call this model $\Lambda$CDM only in the rest of the paper. The second model has four massive neutrinos but they have 
different degeneracy factors and self interaction in sterile sector as described in the earlier sections. 
Mass of the fourth massive neutrino is considered to be 0.2 eV which is the bestfit value of MiniBoone\cite{Aguilar-Arevalo:2018gpe}.
We will refer to this model as SI$\nu$ model.

Self interacting sterile neutrinos affect CMB in three different ways. First, the mass of the sterile neutrino effects the CMB similarly 
the way massive neutrinos effect. It shifts the peaks of the CMB towards the lower values of $\ell$ and decreases the height of the peaks too.  

The second factor which comes into the play is the degeneracy of the mass eigenstates.  
The degeneracy of these mass eigenstates can be written as 
\begin{eqnarray}
 \rho_i = \sum_\alpha|U_{\alpha i}|^2 \rho_{\alpha \alpha}
\end{eqnarray}
For the parameters $U_{\alpha i}$ defined in the earlier sections and the values of $\rho_{\alpha\alpha}$ plotted in \fig{fig:qke} the degeneracy of the corresponding mass eigenstates are given in table~\ref{tab:degen}. 

\begin{table}[h]
\begin{center}
\begin{tabular} {| c | c | c | c | c |}
\hline
Mass ($m_i$) in eV & 0.00001 & 0.009 & 0.05 & 0.2  \\
\hline
Degeneracy($\rho_i$)& 0.69 & 0.84 & 0.90 & 0.56  \\
\hline
\end{tabular}
\end{center}
\caption{Values of degeneracy obtained from QKE for different mass eigenstates 
in SI$\nu$ model}\label{tab:degen}
\end{table}
In \fig{fig:cmb} we have plotted  CMB temperature ($TT$) power spectra for 
$\Lambda$CDM model and then that with the values of neutrino masses and 
degeneracy factors of table~\ref{tab:degen}. { The peaks of CMB 
shifts towards the lower values of $\ell$ in $G_X =0$ case. It is shown in the 
zoomed inset panel of the plot~(a) in \fig{fig:cmb}. However, the effect 
of a 0.2 eV 
massive neutrino becomes milder due to its degeneracy factor of 0.56. It is 
because the fraction of neutrino energy density to the critical density of 
universe ($\Omega_\nu$) depends on the mass of neutrinos and their degeneracies 
in the following way, \cite{cambnotes}\footnote{CAMB and CLASS, both of these 
boltzman codes take care of the effects of degeneracies of neutrinos in similar 
way.} 
\begin{eqnarray}
 \Omega_\nu h^2 = {\sum_i \rho_i^{3/4} m_i\over 93.2 {\rm eV}}\, .
\end{eqnarray}
 This means that a mass eigen state with degeneracy factor less than 
one contributes less in $\Omega_\nu$ as compared to a mass eigen state with 
same mass but degeneracy facotor one. Therefore, in SI$\nu$ model, contribution 
of each state in $\Omega_\nu$ is less as their degeneracy factors are less 
than one. Hence their effect on CMB power spectrum will be milder as compared to the case where 
each state has degeneracy factor one.}

Third effect comes from the self interaction in sterile neutrinos. This effect is also similar to the effect of interactions in active 
neutrino sector studied in literature~\cite{Kreisch:2019yzn}. Self-interaction 
helps to grow the perturbations on small scales and therefore the peaks of the 
CMB grows up due to that. { This effect is visible in both the 
plots~(a) and (b) of \fig{fig:cmb}. In panel (a) where the $TT$ power spectrum 
for different cases is plotted, we can see that there is a increase in the 
height of the peaks for $G_X = 10^{10} G_F$. In panel (b), the ratio of $TT$ 
power 
spectrum for $G_X =0$ and $G_X = 10^{10} G_F$ with $\Lambda$CDM was plotted. The 
effect of of self-interaction
is clearly seen to increase the amplitudes of CMB $TT$ power spectrum for all 
values of $\ell$. In these plots the same mass and degeneracy of 
table~\ref{tab:degen} was used for for $G_X =0$ and $G_X = 
10^{10} G_F$ cases.} 

\begin{table}[h]
\begin{center}
\begin{tabular} { | l | c | c |}
\hline
 Parameter &  $\Lambda$CDM & SI$\nu$ model\\
\hline
{\boldmath$\omega_{b }$} & $ 2.212\pm 0.022  \times 10^{-2}  $ & $2.177\pm 0.022 
   \times 10^{-2}$\\
{\boldmath$\omega_{cdm }  $} & $0.1202\pm 0.0021  $ & $ 0.1220\pm 0.0021   $\\

{\boldmath$100\theta_{s }  $} & $1.04185\pm 0.00045   $ & $1.04481\pm 0.00045    
$\\

{\boldmath$n_{s }         $} & $0.9643\pm 0.0057    $ & $0.9361\pm 0.0055        
  $\\

{\boldmath$\tau_{\rm reio }   $} & $0.0522\pm 0.0079   $  & $0.0476\pm 0.0074    
 $\\

{\boldmath$r              $} & $<0.0460 $  & $<0.0383$\\

{\boldmath$H_0 $} & $67.08\pm 0.92     $ & $66.00\pm 0.91              $\\

{\boldmath$A_{s }  $} & $2.091\pm 0.034  \times 10^{-9}   $ & $1.988\pm 0.030  
\times 10^{-9}  $\\

{\boldmath$\sigma_8        $} & $0.8101\pm 0.0092    $  & $0.7769\pm 0.0086   
$\\
\hline
\end{tabular}
\end{center}
\caption{The best fit values with 1-$\sigma$ error for all the parameters  
obtained from the MCMC analyses for both the
models considered are listed here.}\label{tab:param}
\end{table}

Next, we analyse the effect of self interacting sterile neutrinos on 
cosmological parameters in the 
light of Planck observations~\cite{Aghanim:2018eyx}. We use Planck CMB observations \cite{Aghanim:2018eyx} for temperature anisotropy power spectrum over the multipole range 
$\ell \sim 2$-$2500$ and Planck CMB polarization data for the multipole range $\ell \sim 2$-$29$  only. For this purpose we have used the Planck high-$\ell$ and low-$\ell$ likelihood as defined in ref.~\cite{Aghanim:2018eyx,Akrami:2018odb}. We will refer these data combined as Planck data. For doing the Markov Chain Monte Carlo (MCMC) analysis of the parameter space we have used 
Montepython~\cite{Audren:2012wb}. Sterile neutrino parameters were kept fixed at $G_X=10^{10}G_F$, $\theta_{14}=0.67$ and $m_{st}=0.2$eV 
and all seven cosmological parameters were allowed to vary. The seven cosmological parameters are as follows: The
ratio of density of cold dark matter and baryonic matter evaluated today to the critical density multiplied by 
squared of the reduced Hubble parameter is $\omega_{\rm cdm}$ and $\omega_b$ respectively. Acoustic scale of baryon acoustic oscillation
is $\theta_s$. $A_s$ and $n_s$ are the amplitude and the spectral index of the primordial density perturbations respectively. 
Optical depth to the epoch of re-ionization is denoted by $\tau_{\rm reion}$. And tensor-to-scalar ratio is called $r$. Present Hubble value
$H_0$ and the amplitude of matter power spectrum smoothed over 8$h^{-1}{\rm Mpc}$ scale, known as $\sigma_8$, are the derived parameters.
We list down the best-fit values of these cosmological parameters for the 
two different models in table~\ref{tab:param}. More details about the MCMC is given in appendix~\ref{sec:mcmc}

\begin{figure}
\centering
 \includegraphics[width=9cm,height=8cm]{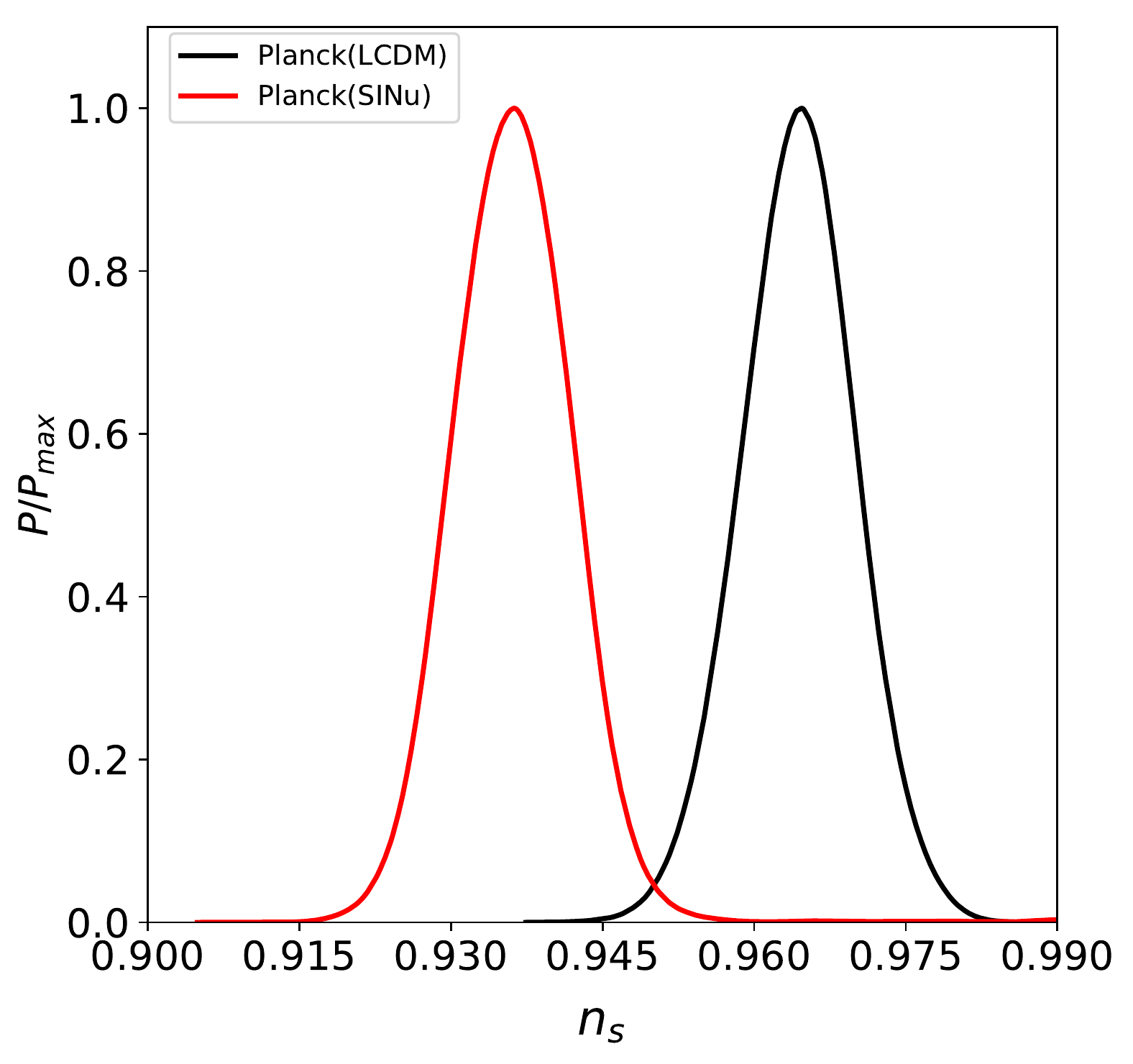}
 \caption{Comparison of the posterior distributions obtained for two models using Planck CMB data.}\label{fig:posterior} 
\end{figure}

We have found that both these models have equivalent statistical significance of fitting Planck CMB data. In \fig{fig:posterior}
the posterior distribution of spectral index has been shown. We see that both the posteriors have same width. Moreover, from table~\ref{tab:param}
we can find out that the one sigma error for $n_s$ is even smaller for the 
SI$\nu$ model. However, there is a increase of 1.45  in the 
maximum value of $-\log($likelihood) in the case of SI$\nu$ model compared to $\Lambda$CDM.

\begin{figure}
\centering
 \includegraphics[width=12cm,height=10cm]{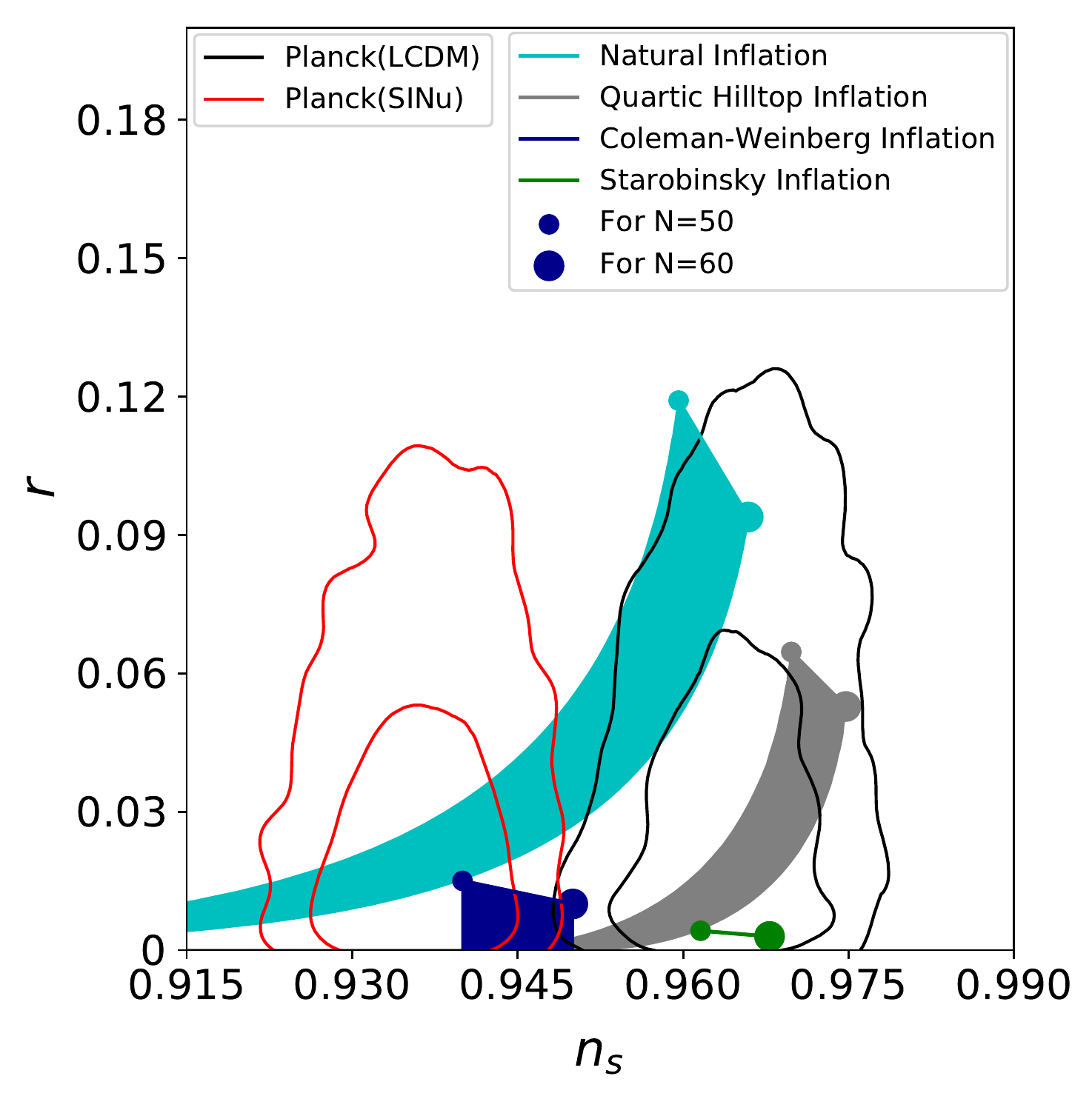}
 \caption{Among the four models shown here only natural inflation satisfies the $n_s$, $r$ values in 1$\sigma$ range. Starobinsky model is
 completely ruled out and Coleman-Weinberg and hilltop inflation fits in the $n_s$, $r$ values for lesser number of $e$-foldings.}
 \label{ns-r}
 \end{figure}

Figures~\ref{ns-r} shows the 1-$\sigma$ and 2-$\sigma$ contours for $n_s$-$r$ planes. The 1-$\sigma$ maximum value of $r$ drops down 
from 0.0460 to 0.0383 for introducing self-interaction in sterile neutrino. In 
last section we have pointed out that 
self-interaction among sterile neutrinos have negligible effect on $BB$ modes of CMB. However, there is an indirect effect on $BB$ spectrum
in this scenario. The increase in the amplitude of $TT$ spectrum in SI$\nu$ model forces other cosmological parameters like $\tau_{\rm reio}$
and $\omega_{\rm cdm}$ to change. This changes effects the $BB$ modes in lower values of $\ell$. To adjust those changes the value of $r$ comes down.
\section{Inflation models}
\label{sec:inf}
In this section we will consider four inflation models. More models can be studied in this context. However, we
find that for these four models our shifts in cosmological parameters make significant implications. The four 
models are Starobinsky inflation, natural inflation, quartic hilltop inflation and Coleman-Weinberg inflation.

Before going to the details of the models, we briefly review the relations of spectral index and tensor-to-scalar 
ratio with the inflaton potential. If the potential of inflation field $\phi$ is denoted by $V(\phi)$ then the 
first order slow roll parameters are written in terms of inflaton potential as
\begin{eqnarray}
 \epsilon &=&\frac{M_{\rm Pl}^2}{2} \left(\frac{V'(\phi )}{V(\phi )}\right)^2 ,\nn\\
 \eta &=& M_{\rm Pl}^2\frac{V''(\phi )}{V(\phi )} .
\end{eqnarray}
Number of $e$-foldings from the horizon exit of the pivot scale to the end of inflation is 
\begin{eqnarray}
 N = {1\over M_{\rm Pl}^2}\int_{\phi_{\rm end}}^{\phi^*} {V(\phi)\over V'(\phi)} \, ,
\end{eqnarray}
where $\phi^*$ is the value of the background inflaton field when pivot scale made the horizon exit, and 
$\phi_{\rm end}$ is at the end of inflation. Therefore, $\phi^*$ is a function of $N$. The value of scalar
spectral index is calculated at the $\phi^*$ and it is 
\begin{eqnarray}
 n_s = 1-6\epsilon(\phi^*) + 2 \eta(\phi^*)\,.
\end{eqnarray}
Similarly tensor-to-scalar ratio is given by
\begin{eqnarray}
 r = 16 \epsilon(\phi^*)\, .
\end{eqnarray}
Amplitude of density perturbation is 
\begin{eqnarray}
 A_s = {1\over 24\pi^2 M_{\rm Pl}^4}{V(\phi*)\over \epsilon(\phi*)}\, .
\end{eqnarray}
With these tools in hand we can calculate the spectral index and tensor-to-scalar ratio from any given inflation
potential.

\subsection{Starobinsky inflation} 
The action for the Starobinsky model is written in Jordan frame as 
\begin{eqnarray}
 S = -{M_{\rm Pl}^2 \over 2 }\int d^4x \sqrt{-g}\left(R +  {R^2\over 6 {\cal M}^2}\right),
\end{eqnarray}
where ${\cal M}$ is a mass scale and $R$ is the Ricci scalar. Upon conformal transformation to Einstein frame
this action reduces to 
\begin{eqnarray}
 S = - {M_{\rm Pl}^2\over 2 }\int d^4x \sqrt{-g}[R-{1\over 2}g^{\mu\nu}\partial_\mu\phi\partial_\nu\phi-V(\phi)],
\end{eqnarray}
where,
\begin{eqnarray}\label{eq:pot-Starobinsky}
 V(\phi) = \Lambda ^4 \left(1-e^{-\sqrt{\frac{2}{3}} \phi }\right)^2,
\end{eqnarray}
and 
\begin{eqnarray}
\phi = \sqrt{3\over 2}M_{\rm Pl}\ln\left( 1+{R\over 3 {\cal M}^2}\right).
\end{eqnarray}
Inflation ends in this model for $\epsilon=1$. From that value of $\phi_{\rm end}$ we follow the procedure
described at the beginning of this section and we come up with a value of $n_s$ to be 0.961585 and 
0.967833 for 50 and 60 $e$-foldings respectively. $\phi^*$ is 5.24$M_{\rm Pl}$ and 5.45$M_{\rm Pl}$ respectively. 
$\eta$ remains negative at these values of $\phi^*$ and $\epsilon$ becomes very small, so we get $r$ to be
0.004 and 0.003 for $N=50$ and 60.

This model of inflation has been the most popular model since the release
of Planck 2013 data where tensor-to-scalar ratio was reported to be low enough to accommodate any convex 
potential. Starobinsky model does not only predict low tensor-to-scalar ratio  but also predicts the value 
of spectral index which fits exactly within 1-$\sigma$ level in $n_s$-$r$ plot for $\Lambda$CDM cosmology(see \fig{ns-r}).
However, the for the case of self-interacting sterile neutrino the prediction of this model lies well outside
the 2-$\sigma$ range of constrained $n_s$-$r$ values. Since the $n_s$ prediction of this model does not depend on any
parameter, this theory cannot be adjusted to fit into the modified  of self interacting sterile neutrino.  

The success of Starobinsky inflation prompted many authors to non-minimally couple a scalar field with the Ricci scalar($R$).
The most popular one is HIggs inflation~\cite{Bezrukov:2007ep} where the scalar field is considered to be standard model 
higgs itself. The action in the Jordan frame has the following form
\begin{eqnarray}
 S = -{M_{\rm Pl}^2 \over 2 }\int d^4x \sqrt{-g}\left[\left(1 +  {\xi\chi^2\over M_{\rm Pl}^2}\right)R-{1\over 2}g^{\mu\nu}\partial_\mu\chi\partial_\nu\chi-V(\chi)\right].
\end{eqnarray}
After conformal transformation to the Einstein frame the potential exactly takes the form of \eqn{eq:pot-Starobinsky}. Therefore
the prediction of this model and Starobinsky model in $n_s$-$r$ plane is absolutely same. So, according to the shifts in $n_s$ in 
SI$\nu$ case, this model will also be ruled out. 

\subsection{Natural inflation} Pseudo-Nambu-Goldstone boson were considered to be a suitable candidate to drive inflation
unless the advent of Planck data. This model has a free parameter $f$ which can be adjusted to shift the values of $n_s$
and $r$. Although the predicted spectral index missed the 1-$\sigma$ contour in $n_s$-$r$ plane for $\Lambda$CDM cosmology.
In our case of self-interacting sterile neutrino the model revives again and the predictions for this model comes well
within the 1-$\sigma$ contour in $n_s$-$r$.

The potential of the model is given by~\cite{Freese:1990rb,Freese:2004un}
\begin{eqnarray}
 V(\phi)=\Lambda ^4 \left[\cos \left(\frac{\phi }{f}\right)\pm 1\right].
\end{eqnarray}
$\Lambda$ determines the scale of the inflation and amplitude of density perturbations. $n_s$ and $r$ are independent of
$\Lambda$ but depends on $f$. The predicted region of $n_s$ and $r$ falls within the allowed 1-$\sigma$ range for self-interacting 
sterile neutrino and the allowed range for $f$ in that case is 3.8$M_{\rm Pl}$ to $4.7 M_{\rm Pl}$ 
\subsection{Quartic hilltop inflation}
Quartic hilltop represents a set of all models whose potential can be represented as \cite{Kohri:2007gq}
\begin{eqnarray}
 V(\phi)= \Lambda^4\left[1-\left(\phi\over v\right)^4 + ...\right].
\end{eqnarray}
Naturally the potential written above is incomplete and unbounded from below. Therefore any complete
model will consist of other extra terms in potential.
For this kind of models, which slowroll parameter among $\epsilon$ and $\eta$ will violate slowroll condition and thus
determine the value of $\phi_{\rm end}$ that depends on the other terms 
of the potential which are not included here. Therefore, value of $\phi_{\rm end}$ at the end of inflation 
varies from model to model. However, for the purpose of a generic discussion we compute the value of $\phi_{\rm end}$
with the above mentioned potential only. Similarly like the natural inflation model the amplitude of density 
perturbation depends on $\Lambda$. Spectral index and tensor-to-scalar ratio is just the function of $v$. We vary the 
value of $v$ from 0.01 to 100 $M_{\rm Pl}$ to get the values of $n_s$ and $r$ shown in \fig{ns-r}. For smaller values of 
$v$, $r$ also becomes smaller. For self-interacting sterile neutrino $v$ is preferred to be less than 1  $M_{\rm Pl}$ and 
7 $M_{\rm Pl}$ for $e$-foldings 60 and 50 respectively.
\subsection{Coleman-Weinberg inflation}
Potential of the model is given by~\cite{Barenboim:2013wra}
\begin{eqnarray}
 V(\phi)= \frac{A f^4}{4}+A \phi ^4 \left[\log \left(\frac{\phi }{f}\right)-\frac{1}{4}\right].
\end{eqnarray}
This model of inflation is the Coleman-Weinberg correction to the $\phi^4$ potential. The model has two parameters
$A$ and $f$. Although the model works for almost all the values of $f$, since it is the renormalization scale it cannot be above 
Planck mass. Moreover, the region of horizon exit in the potential corresponds to $\phi^* \ll f$ region for any value of $f$.
For $N=50$ and 60 the $n_s$ turns out to be 0.94 and 0.95 respectively. The maximum possible value of $r$ for these values of 
$N$ is 0.015 and 0.01. These values of $r$ can go down to much smaller values depending on the value of $f$.

\section{Discussion and Conclusion}
\label{sec:conclusion}
Constraining inflation models on the basis of spectral index and tensor-to-scalar ratio obtained from the CMB data had been the forefront
area of research in cosmology and high energy physics for last decades. 
Inflation models are constrained from CMB observations which place limits on inflation parameters like amplitude of primordial power spectra, tilt of the scalar power spectra and tensor to scalar ratio. In these analyses the base model is normally the $\Lambda$CDM cosmology. 
Although there are many tensions like $\sigma_8$~\cite{Joudaki:2016kym,Anand:2017wsj,Mohanty:2018ame,Lambiase:2018ows} 
or $H_0$ tension~\cite{Bernal:2016gxb}
among different cosmological measurements under $\Lambda$CDM framework still it is undeniable that $\Lambda$CDM is the best model available
so far to fit the CMB data. The need to accommodate sterile neutrinos, motivated 
by SBL neutrino oscillation experiments, requires one to introduce large self 
interaction in the sterile neutrino sector. We studied the change in 
cosmological parameters in this scenario. We find that the best fit values of 
$n_s=0.9643 \pm 0.0057$ in $\Lambda$CDM model changes to $n_s= 0.9361\pm 
0.0055$ in SI$\nu$ model. This happens because
self-interaction in sterile neutrino shifts the CMB peaks in upward. 
Although there are many theoretical models which, when examined with CMB data, suggests  different values 
of spectral index and tensor-to-scalar ratio, but most of them either use extra 
cosmological parameters or they decrease the value of likelihood. 
Interestingly SI$\nu$ model, with the neutrino parameters fixed from MiniBooNE, achieves the similar likelihood values as the $\Lambda$CDM model with no other extra free cosmological parameter.

The change in inflation parameters make significant impact on the validity of different cosmological models.
In this paper we mainly focused on those potentials which have a concave section
since self-interacting sterile neutrino did not enhance the tensor-to-scalar ratio.
We found that the Starobinsky model which is the best fit model in $\Lambda$CDM cosmology is ruled out in the SI$\nu$ cosmology. 
The similar models which use non-minimal coupling of scalar field to gravity sector for achieving inflation is ruled out too.
In addition the quartic hilltop is also disfavored in the SI$\nu$ model which was allowed at 1-$\sigma$ level in $\Lambda$CDM model. On the other hand some models like natural and Coleman-Weinberg inflation, which were ruled out in the $\Lambda$CDM cosmology are now favored. 
In conclusion if future SBL neutrino oscillation experiments confirms the existence of eV scale sterile neutrino then it will have significant impact on the choice of viable inflation models.

\section* {Appendix}
\appendix
\section{Details of MCMC}\label{sec:mcmc}
The priors of the seven cosmological parameters are given in 
table~\ref{tab:prior}. We have used Gaussian prior for our purpose. In case 
of $r$ a maximum and a minimum value was assigned which were 0.5 and 0.0 
respectively. Similarly a minimum value of $\tau_{\rm reio}$ was specified
to be $0.004$.

\begin{table}[h]
\begin{center}
\begin{tabular} { | l | c | c |}
\hline
Parameter & mean & 1-$\sigma$\\
\hline
$\omega_b$ & $2.2377\times 10^{-2}$ &  $0.015\times 10^{-2}$\\
$\omega_{\rm cdm}$  & 0.12010 & 0.0013 \\
$100\theta_s$  & 1.04110  & 3e-4 \\
$\ln 10^{10}A_s$ & 3.0447 & 0.015 \\
$n_s$          & 0.9659 & 0.0042 \\
$\tau_{\rm reio}$   & 0.0543 &   0.008\\ 
$r$            & 0.06 & 0.04\\
\hline
\end{tabular}
\end{center}
\caption{Priors used in MCMC}\label{tab:prior}
\end{table}

\begin{figure}[h]
 \includegraphics[width=19cm,height=15cm]{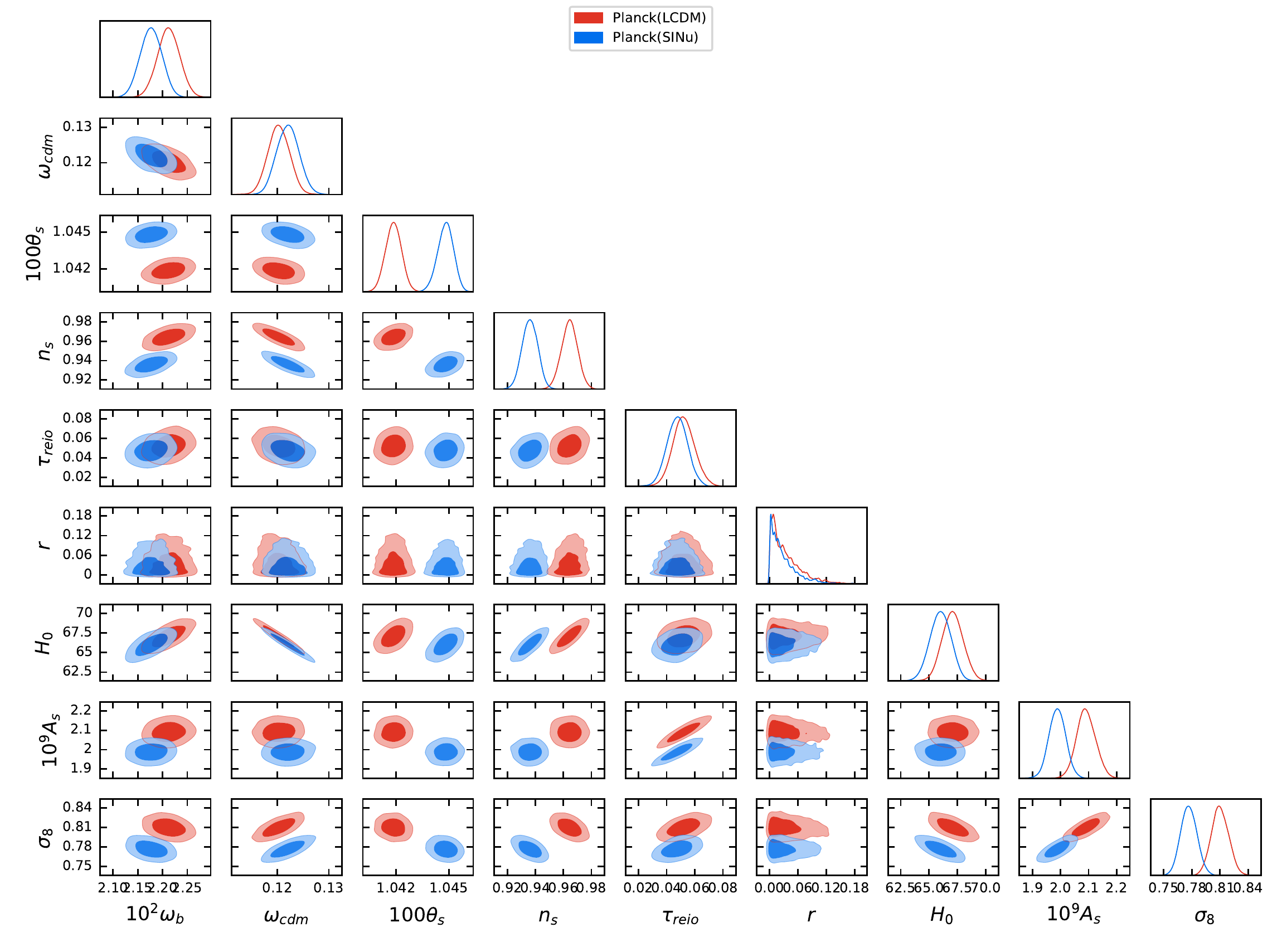} 
\caption{Triangle plot of the posterior distributions for the cosmological 
parameters.}\label{fig:triangle}
 \end{figure}

In \fig{fig:triangle} we can see that two parameters gets affected significantly and these two parameters are spectral index $n_s$ and 
acoustic scale $\theta_{s}$. Since $H_0$ is the derived parameter from $\theta_s$ the value of $H_0$ also decreases making the $H_0$ tension
even more severe. The amplitude of matter power spectrum smoothed over 8$h^{-1}{\rm Mpc}$ scale, known as $\sigma_8$, also reduces because 
of the effect of massive neutrino on matter power spectrum~\cite{Anand:2017ktp}.

We can see from these posterior plots in \fig{fig:triangle} that the posteriors has almost same $\sigma$ for both the models. This makes the two models
equivalent in fitting the Planck CMB data.

\section* {Acknowledgment}
We would like to thank Ningqiang Song for suggestions in solving QKE. We would also like to thank Akhilesh Nautiyal
for various helps regarding plotting. We also acknowledge the computation facility, 100TFLOP HPC Cluster, Vikram-100, at Physical Research Laboratory, Ahmedabad, India.

\newpage

\bibliographystyle{unsrt}%
\bibliography{paper_draft.bib}  
  
\end{document}